\documentclass[]{spie}

\usepackage{amsmath,amsfonts,amssymb}
\usepackage{graphicx}
\usepackage[version=4]{mhchem}
\usepackage[colorlinks=true, allcolors=blue]{hyperref}
\usepackage{siunitx}

\newcommand{\bettersim}{{\raise.17ex\hbox{$\scriptstyle\sim$}}}

\DeclareSIUnit{\sample}{S}
\DeclareSIUnit{\photon}{ph}
\DeclareSIUnit{\electron}{\mathrm{e^-}}
\DeclareSIUnit\bar{bar}

\title{The Integrating Miniature Piggyback for Impulsive Solar Hard X-rays (IMPISH): a spectrometer for the GRIPS-2 balloon campaign}

\author[a]{Reed B. Masek}
\author[a]{William Setterberg}
\author[a]{Dorcas Oseni}
\author[a]{Lestat Clemmer}
\author[a]{Lindsay Glesener}
\author[a]{Philip Williams}
\author[b]{John G. Sample}
\author[c]{Amir Caspi}
\author[a]{Demoz Gebre-Egziabher}
\author[d]{Pascal Saint-Hilaire}
\author[e]{Albert Y. Shih}
\author[f]{David M. Smith}
\affil[a]{University of Minnesota, Minneapolis, USA}
\affil[b]{Montana State University, Bozeman, USA}
\affil[c]{Southwest Research Institute, Boulder, USA}
\affil[d]{University of California, Berkeley, Berkeley, USA}
\affil[e]{NASA Goddard Space Flight Center, Greenbelt, USA}
\affil[f]{University of California, Santa Cruz, Santa Cruz, USA}

\authorinfo{Additional author information: (Send correspondence to R.B.M.)\\R.B.M.: E-mail: masek014@umn.edu\\}

\begin{document} 
\maketitle

\begin{abstract}
   The Integrating Miniature Piggyback for Impulsive Solar Hard X-rays (IMPISH) is a piggyback mission originally designed for the second flight of the Gamma-Ray Imager/Polarimeter for Solar flares (GRIPS-2) Antarctic balloon. IMPISH will take measurements of collimated, full-Sun X-ray spectra with the goal of detecting sub-second variations (order of tens of milliseconds) of nonthermal X-ray emission during the impulsive phase of large solar flares to probe particle acceleration mechanisms driven by magnetic reconnection.

   The IMPISH detector system, made up of four identical detectors totaling \SI{64}{\centi\meter\squared} effective area, is capable of measuring from \SI{\bettersim 10}{\kilo\eV} to over \SI{200}{\kilo\eV} through the use of silicon photomultipliers (SiPMs) and LYSO scintillators. At the stratospheric altitude of GRIPS-2, the effective lower energy limit is \SIrange{\bettersim 20}{30}{\kilo\eV}. The geometry of the LYSO crystal has been optimized to balance the light collection efficiency with the effective area required for stratospheric X-ray measurements. Development of the IMPISH detectors has introduced a path for a low-cost solution to fast solar X-ray measurements across a large energy range utilizing commercially available components. The payload has a 3U form factor and has been designed so that both the electronics and detectors may be easily adaptable for space-based missions.
\end{abstract}


\section{INTRODUCTION}%
\label{sec:introduction}

\begin{figure*}
   \centering
   \begin{minipage}{.5\textwidth}
      \centering
      \begin{tabular}{c}
         \includegraphics[width=\linewidth]{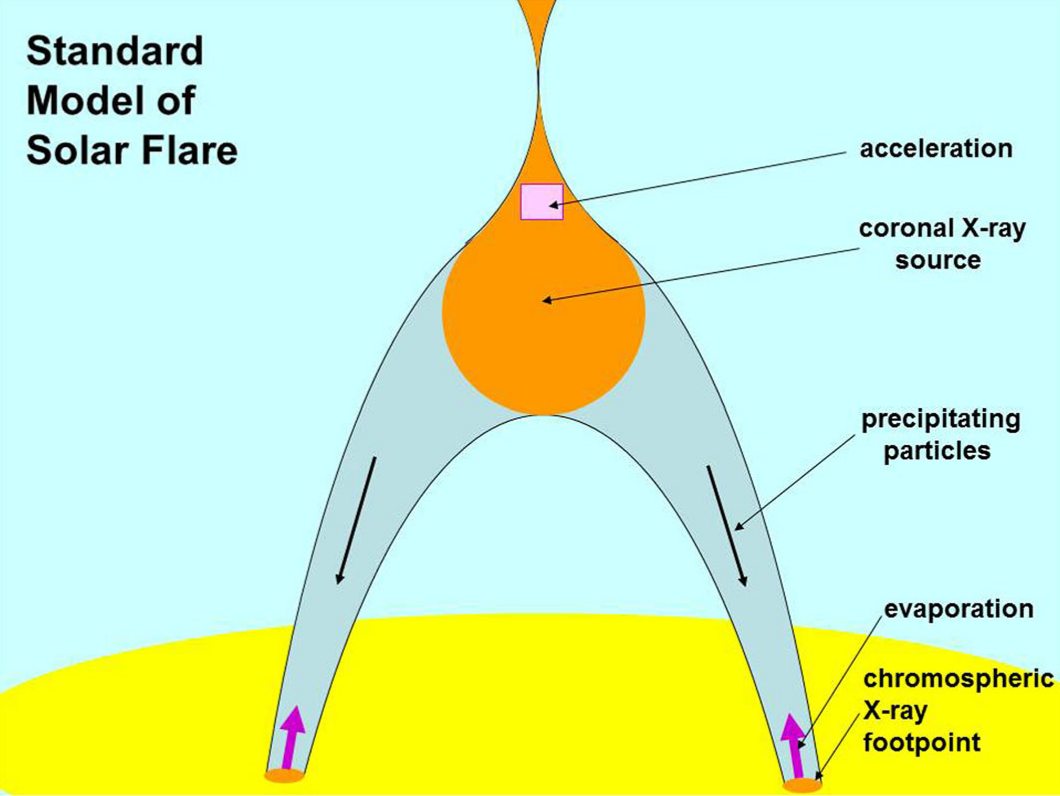}
      \end{tabular}
   \end{minipage}%
   \begin{minipage}{.5\textwidth}
      \centering
      \begin{tabular}{c}
         \includegraphics[width=0.75\linewidth]{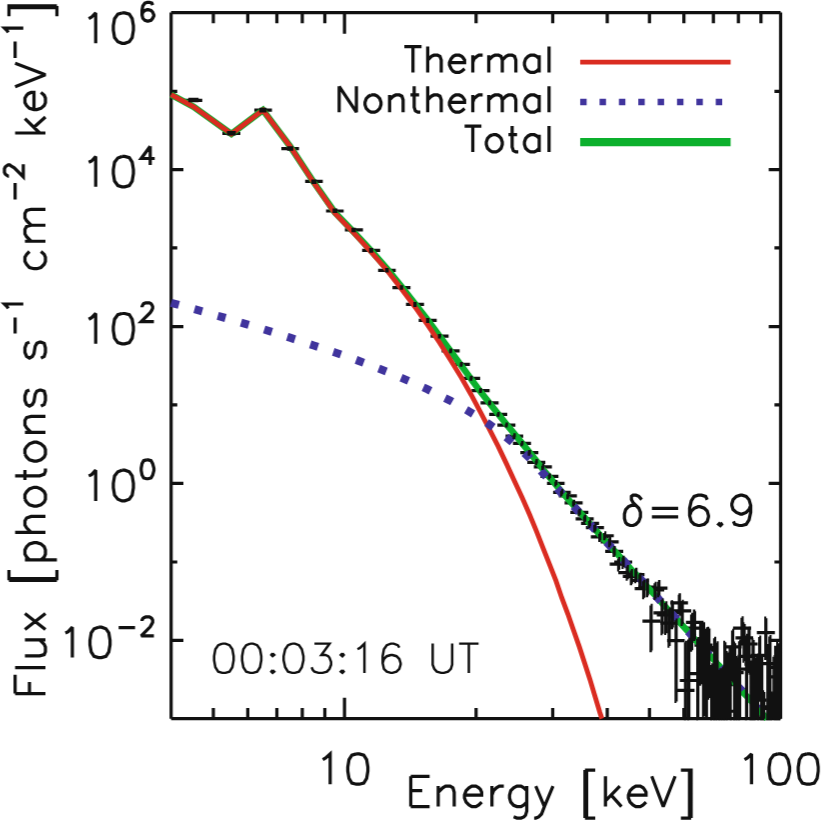}
      \end{tabular}
   \end{minipage}
   \caption
   {\label{fig:flare-cartoon-and-spectrum} \textit{Left}: the standard model for a solar flare, depicting the relevant locations of observed X-ray emission, from Benz (2017).\cite{Benz-2017} Magnetic reconnection occurs high in the corona, accelerating electrons (precipitating particles) toward the solar surface that then interact and produce the X-ray footpoints. \textit{Right}: example of an X-ray spectrum from a flare measured by RHESSI, showing the thermal and nonthermal emission, from Veronig \& Brown 2004.\cite{Veronig-2004}}
\end{figure*}

A solar flare is a violent release of energy stored in the Sun's magnetic field.
Flares heat coronal plasma to temperatures in excess of tens of millions of degrees and emit electromagnetic radiation in all wavelengths.
Different mechanisms are associated with the production of different bandwidths of radiation.
For example, hard X-rays (HXRs), defined from a few to several hundreds of \si{keV}, are produced by processes most closely linked to the initial energy release and subsequent heating.
``Thermal'' emission, or free-free emission, is bremsstrahlung from a hot plasma as a result of a free electron Coulomb scattering with an ion.\cite{Benz-2017}
X-ray energies from these interactions range from sub-\si{\kilo\eV} up to tens of \si{\kilo\eV} and, along with atomic emission lines, can be used for measuring plasma temperatures.
Flare-accelerated electrons from high in the solar corona travel towards the solar surface, interact and impart their energy to the lower-altitude plasma (e.g. in the chromosphere), and generate higher energy X-rays referred to as ``nonthermal'' emission.\cite{Aschwanden-2006}
Nonthermal X-rays energies range from below \SI{10}{\kilo\eV} to several hundreds of \si{\kilo\eV} and are used to probe the acceleration mechanisms driven by the energy released by solar flares.\cite{Glesener-2020, Benz-2017}
As X-ray energy increases, nonthermal emission dominates over thermal emission, so higher energy nonthermal emission has relatively lower background than lower energy emissions.
A depiction of the standard flare model and an example of an X-ray spectrum are shown in Figure~\ref{fig:flare-cartoon-and-spectrum}, showing the nonthermal footpoint and coronal emissions.
Thermal X-rays are emitted as hot plasma fills the loop following the energy release.

A variety of mechanisms have been proposed to be responsible for electron acceleration in solar flares; for example, DC electric field acceleration, stochastic acceleration (random energy gains/losses in a turbulent plasma resulting in a net energy gain), shock acceleration (i.e. first-order Fermi acceleration).\cite{Aschwanden-2006}
Measuring the duration, periodicity, and timing difference across energy bands of nonthermal HXR emission probes the nature of these acceleration mechanisms.\cite{Aschwanden-1996}
For example, prior studies have found sub-second variations in nonthermal HXR emission in solar flares.\cite{Aschwanden-1996,Knuth-2020}
This information can be used with spectroscopic techniques assuming thick- or thin-target models to extract information directly related to the distribution and energetics of the flare-accelerated electron population.\cite{Brown-1971}

Previous solar-dedicated HXR instruments have been incapable of the fast measurement cadence necessary to study these sub-second variations.
The \textit{Reuven Ramaty High Energy Solar Spectroscopic Imager} (RHESSI) was capable of achieivng high energy and spatial resolution of HXRs and gamma rays from solar flares; however, its indirect imaging method made use of a Fourier transform technique that required the entire spacecraft to rotate with a \SI{4}{\second} period, making it very difficult or impossible to use data finer at intervals less than \SI{2}{\second}.\cite{Lin-2002}
The currently operating Spectrometer/Telescope for Imaging X-rays (STIX) aboard Solar Orbiter also uses an indirect imaging technique, but is able to achieve faster cadences than RHESSI.\cite{Krucker-2020}
STIX is capable of sampling down to \SI{0.1}{\second} and has been shown to detect variations on the order of seconds.\cite{Collier-2023}

This paper introduces the Integrating Miniature Piggyback for Impulsive Solar Hard X-rays (IMPISH), which is a low-cost spectrometer for measuring fast variations in HXRs from solar flares with a \SI{\bettersim 30}{\milli\second} sample cadence.
IMPISH was originally designed as a piggyback payload on the second Gamma Ray Imager/Polarimeter for Solar Flares (GRIPS-2) Antarctic balloon mission, although its compact size and low accommodation needs make it suitable for any stratospheric or space based platform.
Section~\ref{sec:impish} details the science goals, mission parameters, and atmospheric environment, while Section~\ref{sec:instrument-design} goes into our crystal and detector designs and custom electronics.
Sections~\ref{sec:current-investigations} and~\ref{sec:conclusions} describe our current investigations and conclusions, respectively.

\section{IMPISH Science}%
\label{sec:impish}

IMPISH is a full-Sun (non-imaging) HXR spectrometer designed for a stratospheric balloon but could be easily adapted to other space-based or near-space platforms.
Current parameters are designed for a piggyback balloon launch from McMurdo Station, Antarctica for a flight lasting one to several weeks.

The ongoing IMPISH payload design and development is led by the University of Minnesota (UMN) in collaboration with Montana State University (MSU), the Southwest Research Institute (SwRI), and the University of California, Santa Cruz (UCSC).
Its primary mission objectives are:
\begin{itemize}
   \item \textbf{Investigate electron acceleration timescales in solar flares.} Sub-second variations in HXR time profiles are a probe into how flares accelerate electrons. Such variations are understudied as many HXR instruments either use indirect imaging techniques involving rotating collimators, e.g. RHESSI, or are subject to intense pileup for moderate-sized flares, e.g. the Fermi Gamma-ray Burst Monitor (GBM). IMPISH will achieve a \SI{32}{\hertz} sample rate, higher than previous solar-dedicated, state of the art spectrometers and will be impervious to pileup except for intense, high X-class flares.
   \item \textbf{Demonstrate a low-resource, fast, high-energy HXR detector.} Solar flare monitoring is becoming increasingly relevant with the rise of space weather prediction and nowcasting. Most space-based solar HXR instrumentation is designed with the purpose of more complex science objectives, resulting in a lack of low-latency solar HXR data, e.g. NOAA's Geostationary Operational Environmental Satellite (GOES) suite. HXRs can give early alerts about big flares, making it desirable to have constant HXR monitoring with quickly accessible data. IMPISH is designed solely for the purpose of performing fast measurements of solar X-rays making use of off-the-shelf components and other commercially available components and could be easily adapted to a space-based environment.
\end{itemize}

The detectors developed for IMPISH seek to improve upon those from a prior design for the Impulsive Phase Rapid Energetic Solar Spectrometer (IMPRESS) CubeSat.\cite{Knuth-Thesis,Setterberg-2022}
The two missions share the same $3\sigma$ criterion for observing statistically significant sub-second spikes in electron time variations.\cite{Knuth-2020,Knuth-Thesis}
However, IMPISH is using a non-hygroscopic scintillator crystal, making it easier to work with in comparison to the hermetically-sealed cerium bromide crystals used on IMPRESS.
This has enabled investigations into alternative reflectors, surface polishing, crystal shapes, and means of optical coupling, which is explored and detailed in a companion paper in this conference (Oseni et al. 2025).\cite{Oseni-2025}

\begin{figure}[t]
   \begin{center}
      \begin{tabular}{c}
         \includegraphics[width=\linewidth]{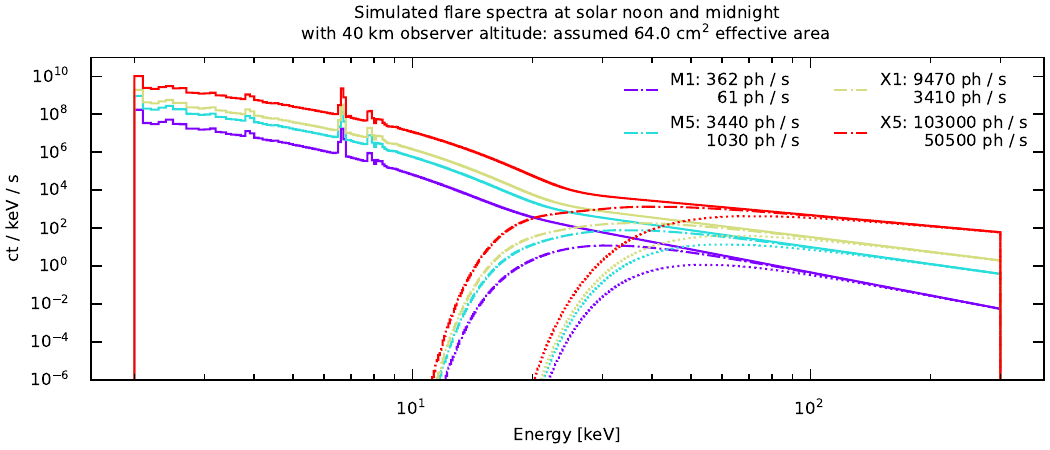}
      \end{tabular}
	\end{center}
   \caption
   {\label{fig:atmospheric-attenuation} Simulated spectra from various magnitudes of solar flares (solid lines) as observed at \SI{40}{\kilo\meter} altitude above McMurdo Station, Antarctica at solar noon (dash-dotted lines) and solar midnight (dotted lines). The photon rates displayed at the top right are computed for a \SI{64}{\centi\meter\squared} effective area (assumed constant across X-ray energies) integrated across the entire energy range. The upper number is computed at solar noon (\SI{54.8}{\degree} zenith near 14:00) while the lower number is at solar midnight (\SI{89.0}{\degree} zenith near 02:00)}
\end{figure}

\subsection{Observational Prospects}%
\label{subsec:observational-prospects}

A balloon launched during the Antarctic summer has a constant view of the Sun throughout the flight.
The balloon's float altitude changes diurnally by several kilometers and, in the case of GRIPS-1, reached a maximum altitude of \SI{40}{\kilo\meter}.\cite{Duncan-2016}
Lower energy X-rays will be attenuated depending on the altitude of the balloon and the zenith angle of the Sun (the angle defined relative from the ground normal), which was characterized through basic simulations.

We used the NRLMSIS 2.0 emperical model, specifically the \texttt{pymsis} wrapper, to obtain estimates of the atmospheric composition at the time and location of our observations \cite{Emmert-2021,Lucas-2023}.
This data was used in conjunction with mass attenuation coefficients to propagate simulated solar flare X-ray spectra from the top of the atmosphere down to \SI{40}{\kilo\meter} altitude.\cite{Hubbell-2004}
Spectra from a range of flare classes, from M1 to X5, were generated using scaling relations and used as inputs to our model.\cite{Battaglia-2005,Knuth-Thesis}
Figure~\ref{fig:atmospheric-attenuation} compares the original, simulated spectrum (at the top of the atmosphere) to two propagated down to \SI{40}{\kilo\meter} (one for solar noon and one for solar midnight) for each of the flare classes.
We used these results to evaluate the effective area needed to achieve our mission goals.
At float altitude, we can see that the atmosphere will greatly attenuate solar X-rays below \SI{40}{\kilo\eV} and completely block those near and below \SI{10}{\kilo\eV}.
These simulations do \textit{not} account for Compton scattering due to the atmosphere, which will likely reduce the flux reaching our detector.

In order to confidently detect an X-ray spike at the $3\sigma$ level assuming a \SI{64}{\centi\meter\squared} effective area and \SI{32}{\hertz} frame rate, we would require something on the order of \SI{5000}{\photon\per\second} detected across any energy range.
As shown in Figure~\ref{fig:atmospheric-attenuation}, we cross this threshold somewhere between an M5-class flare and X1-class flare at solar noon.
Additionally, flares of the same class can greatly vary from one another, so these models are not definitive.
For a 4 week-long flight in December 2026/January 2027, we expect approximately \numrange{5}{10} M-class flares and \num{1} X-class flare based on the solar cycle.\cite{Knuth-Thesis}
However, the number of flares we observe is entirely dependent on solar activity during our flight, which can vary wildly even near solar maximum.

\section{Instrument Design}%
\label{sec:instrument-design}

\begin{figure}[t]
   \begin{center}
      \begin{tabular}{c}
         \includegraphics[width=0.75\linewidth]{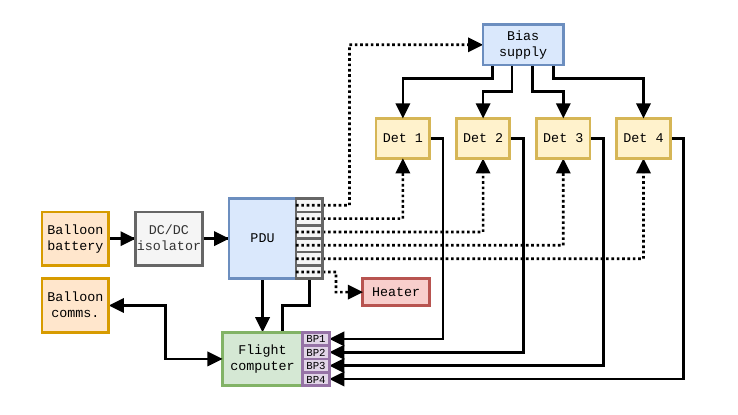}
      \end{tabular}
	\end{center}
   \caption
   {\label{fig:electrical-block} Block diagram of our subsystems. The power distribution unit (PDU) is a set of two boards that supplies all subsystems with power, and the connected dotted lines denote independently toggleable power connections. Each detector (Det) output has its own ADC (Bridgeport SiPM-3000; BP) that digitizes the SiPM pulses and is connected to the flight computer via USB.}
\end{figure}

The IMPISH payload approximately follows a 3U (1U is \numproduct{10 x 10 x 10} \si{\centi\meter\cubed}) CubeSat form.
It is designed to accept an unregulated battery voltage ranging \SIrange{20}{30}{\volt} that is fed into several switching supplies for powering its subsystems.
Many subsystems required by CubeSats are not needed for IMPISH since the attitude, GPS, radio, etc. are handled by the balloon gondola.
Consequentially, IMPISH consists only of power distribution boards, a flight computer (Raspberry Pi Compute Module \num{4}), heaters, and four detector systems.
In the case IMPISH flies on an instrument that is \textit{not} (anti-)Sun-pointed, an aspect system will be developed to point our detectors at the Sun.
The four identical detector systems can be toggled on or off independently of each other and are controlled by the flight computer.
Additionally, the detector bias supply and payload heaters can be toggled or adjusted by the flight computer.
A block diagram of the electronics is shown in Figure~\ref{fig:electrical-block}.

\begin{figure}[t]
   \begin{center}
      \begin{tabular}{c}
         \includegraphics[width=0.55\linewidth]{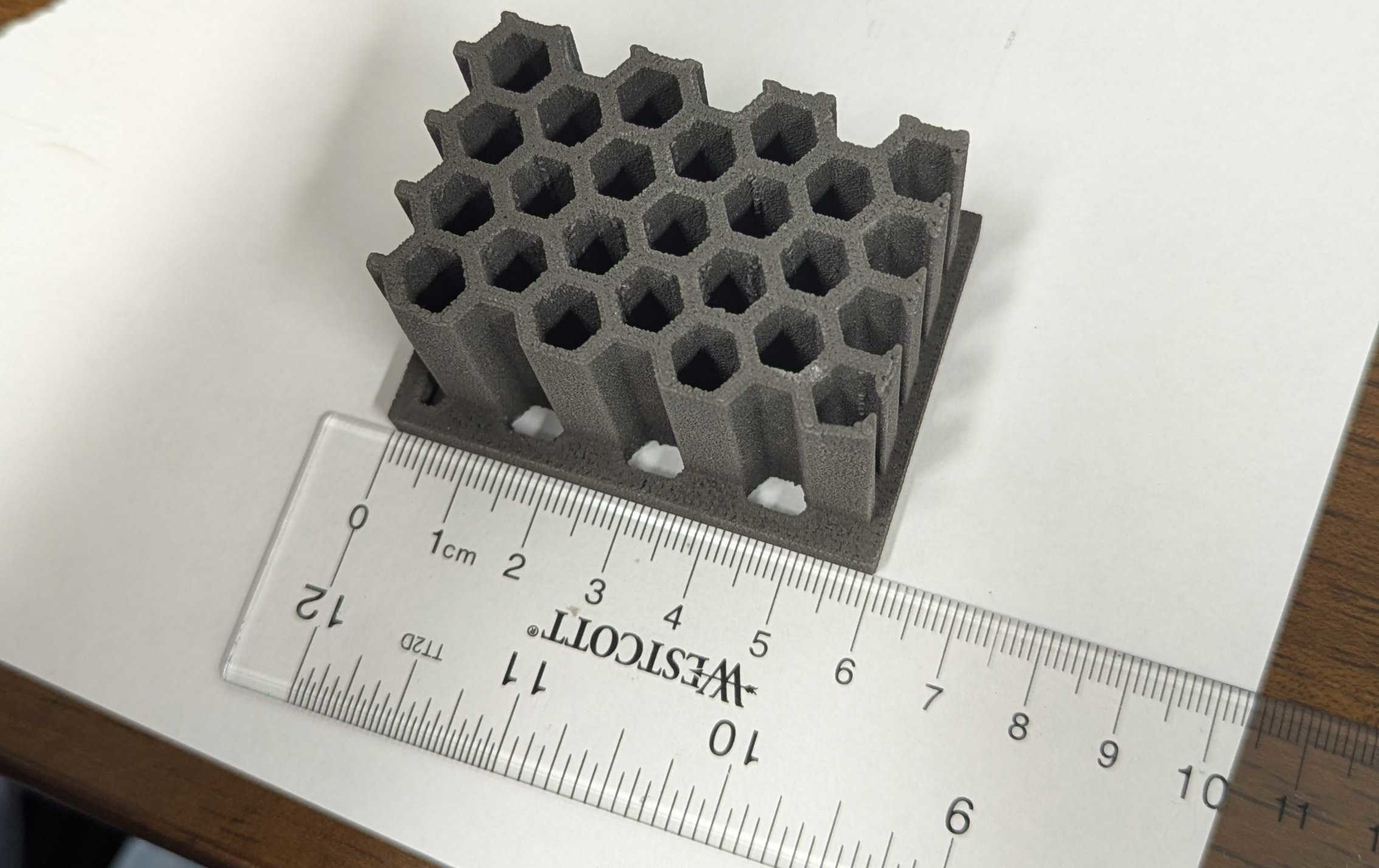}
      \end{tabular}
	\end{center}
   \caption
   {\label{fig:collimator} A sample collimator printed with the tungsten filament. The length and width is scaled down, and the walls are thicker than what will be used in the final version. We are using this sample print to understand the properties of the filament for use with our collimator.}
\end{figure}

\begin{figure*}[t]
   \centering
   \begin{minipage}{.5\textwidth}
      \centering
      \begin{tabular}{c}
         \includegraphics[width=\linewidth]{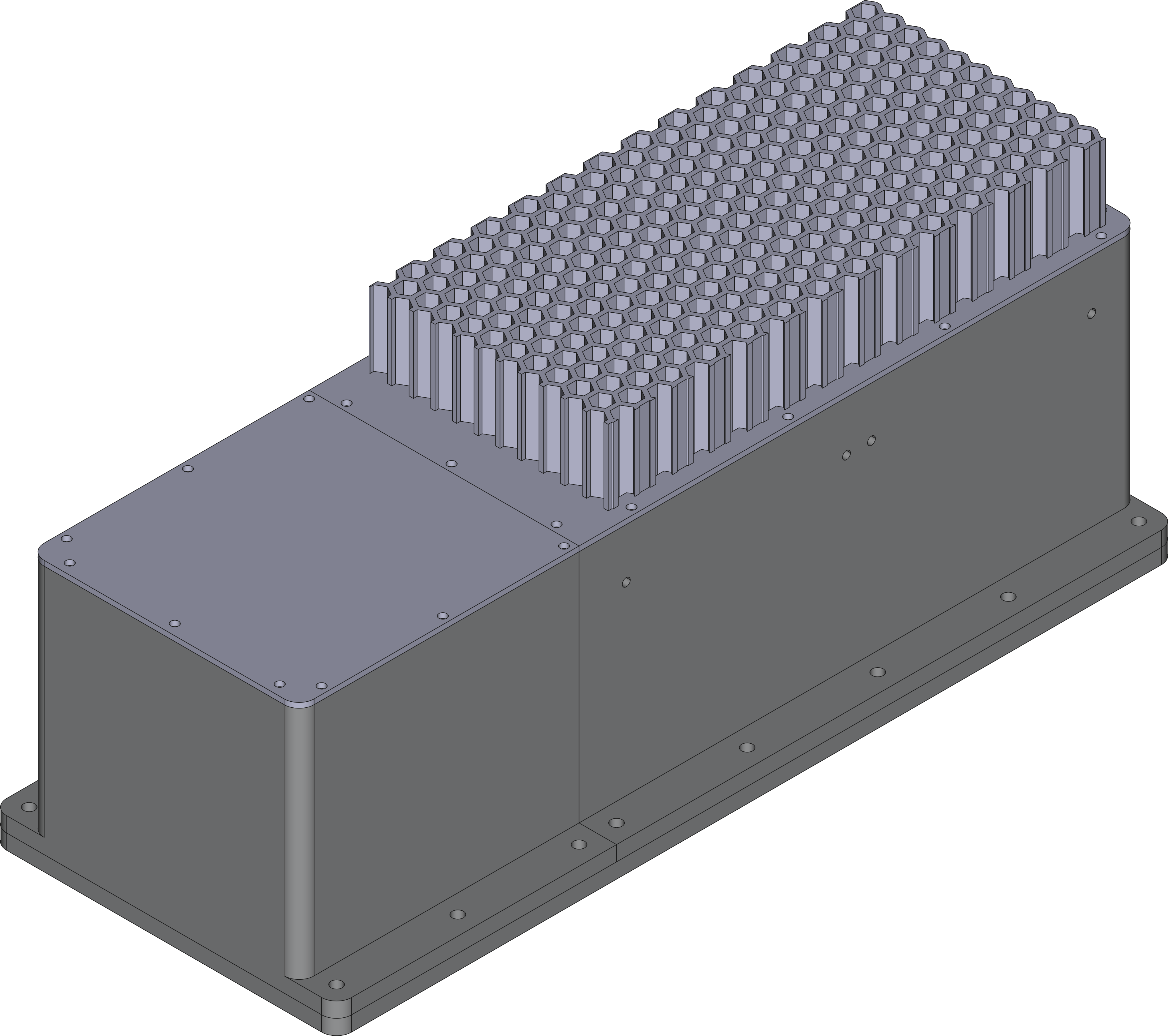}
      \end{tabular}
   \end{minipage}%
   \begin{minipage}{.5\textwidth}
      \centering
      \begin{tabular}{c}
         \includegraphics[width=0.5\linewidth]{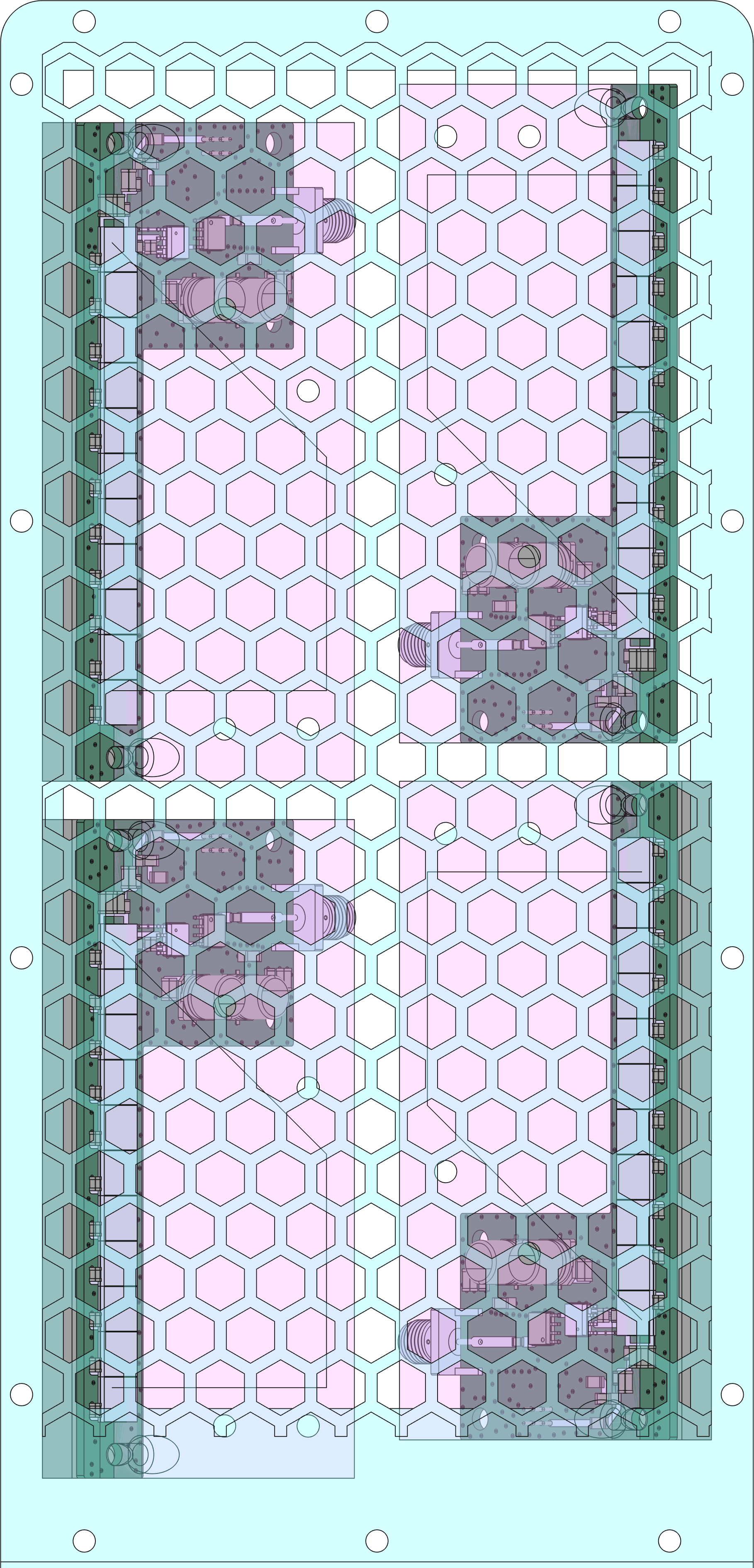}
      \end{tabular}
   \end{minipage}
   \caption
   {\label{fig:payload} \textit{Left}: CAD model of the whole IMPISH payload, approximately 3U in size. The tungsten-based collimator can be seen as the hexagonal grid atop the payload. \textit{Right}: top-down CAD view of the light tight detector box. The light blue, hexagonal grid is the honeycomb collimator, the SiPM PCBs are shown in green, and the scintillator mounts are shown in pink.}
\end{figure*}

The bulk structure will be made of \num{6061}-T6 aluminum alloy, which is commonly used for CubeSats.
All four detector units will be housed inside a light tight box with a single tungsten-based collimator mounted on the sunward face to reduce flux from X-rays scattered by the atmosphere.
One detector unit consists of a scintillator crystal coupled to a silicon photomultiplier (SiPM) array.
A CAD model of a preliminary payload design along with a top-down view of the detector portion (collimator covering the four detector units) can be seen in Figure~\ref{fig:payload}.
The output of a single SiPM array is fed into its own SiPM-3000, a \SI{40}{\mega\sample\per\second} multichannel analyzer from Bridgeport Instruments.
A stock SiPM-3000 has a built-in bias supply and preamplifier, but we have opted to replace them in favor of custom designs in order for more control over the noise performance.
The built-in FPGA allows us to finely tune the integration parameters for use with our specific detector.
It measures the energy by summing the pulse values and subtracting a baseline, and it is capable of producing energy histograms at a cadence of \SI{32}{\hertz}, enabling our science goal of measuring fast (subsecond) nonthermal HXR variations.
For a stratospheric balloon flight, we expect about \numrange{5}{10} \si{\mega\byte\per\day} downlink, assuming a total of four hours of flare observation and ten minutes of background measurements each day.
We detail the collimator, detector unit construction, and custom circuitry in the following sections.

\subsection{Tungsten Collimator}%
\label{subsec:tunsten-collimator}
In order to reduce the X-ray flux scattered by the atmosphere, we have designed a collimator that gives IMPISH a \SI{30}{\degree} field of view with 75\% open area.
The design consists of a honeycomb structure with a wall thickness of \SI{2}{\milli\meter}, equivalent to the mean free path of photoelectric absorption at \SI{200}{\kilo\eV}.
For the material, we chose a tungsten-based filament that is 92\% tungsten by weight and can be 3D printed similar to regular PLA filament with the use of a hardened steel nozzle.
We printed a test cube and measured the resulting density to be approximately \SI{6.6}{\gram\per\centi\meter\cubed}, which is within the specifications reported by the manufacturer.
We have also printed a scaled down model of our collimator, shown in Figure~\ref{fig:collimator}, for testing its effectiveness at blocking X-rays at our desired energies and understanding the mechanical properties of the material.
The necessity of a collimator is enforced by the stratospheric environment, so a space-based adaptation of IMPISH would not require it.

\subsection{Detector Design}%
\label{subsec:detector-design}

\begin{figure}[t]
   \begin{center}
      \begin{tabular}{c}
         \includegraphics[width=\linewidth]{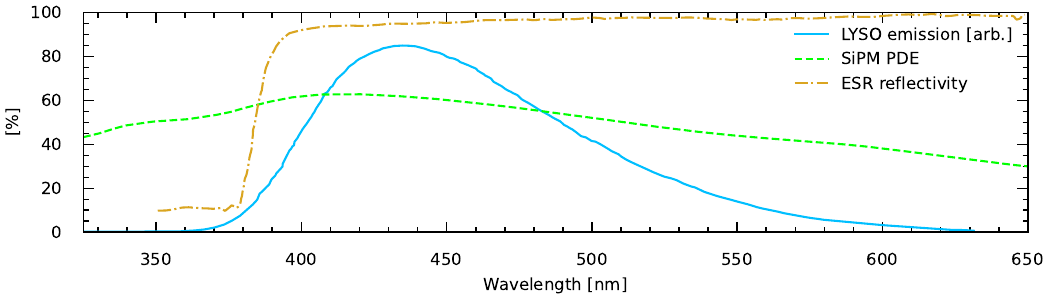}
      \end{tabular}
	\end{center}
   \caption
   {\label{fig:lyso-esr-sipm} LYSO emission spectrum overplotted with ESR reflectivity (at \SI{15}{\degree} incidence angle) and the PDE of our SiPMs (Broadcom AFBR-S4N66P014M). The LYSO emission is plotted in arbitrary units, the ESR reflectivity is plotted as the percent chance the incident photon reflects, and the SiPM PDE is plotted as the percent chance a photon at a particular wavelength is detected by the SiPM (i.e. triggers an avalanche).\cite{Lorincz-2010}}
\end{figure}

We chose LYSO:Ce (cerium-doped lutetium yttrium orthosilicate; \ce{Lu2SiO5}:Ce; hereafter abbreviated to LYSO) for our scintillator.
A detailed analysis of the investigations into this choice is presenting an accompanying paper, Oseni et al. (2025).\cite{Oseni-2025}
Its properties are generally favorable for our mission, namely its fast decay constant of \SI{40}{\nano\second}, relatively bright light yield of about  \SI{30000}{\photon\per\MeV}, non-hygroscopicity, has negligible afterglow, and spectral emission in visible wavelengths (see Figure~\ref{fig:lyso-esr-sipm}) that is well-matched to commonly used photomultipliers.
Its high density of \SI{7.3}{\gram\per\centi\meter\cubed} gives it a high stopping power, allowing the use of thinner crystals.

\begin{figure*}
   \centering
   \begin{minipage}{.5\textwidth}
      \centering
      \begin{tabular}{c}
         \includegraphics[width=\linewidth]{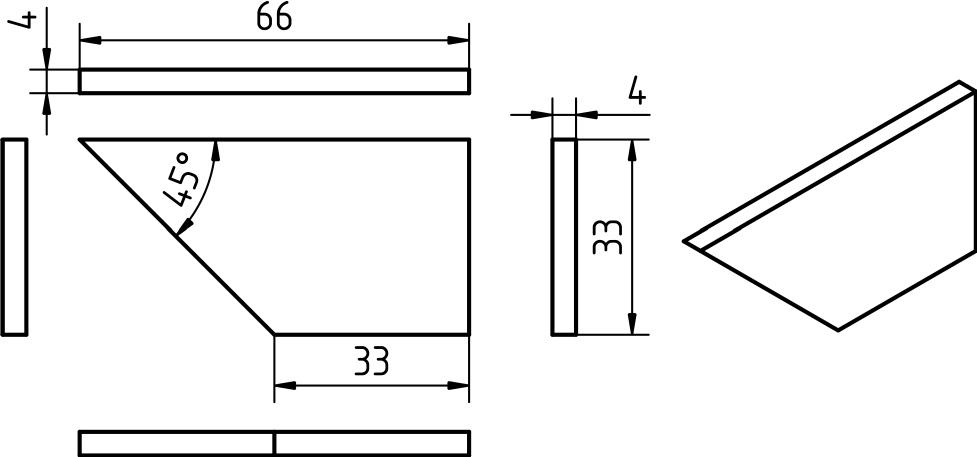}
      \end{tabular}
   \end{minipage}%
   \hfill
   \begin{minipage}{.48\textwidth}
      \centering
      \begin{tabular}{c}
         \includegraphics[width=\linewidth]{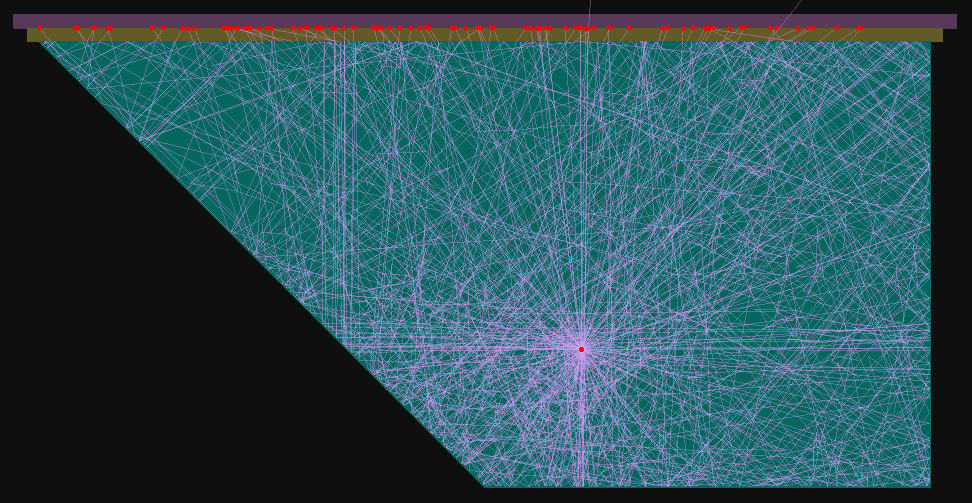}
      \end{tabular}
   \end{minipage}
   \caption
   {\label{fig:crystal-diagram} \textit{Left}: drawing of our selected crystal geometry with dimensions in \si{\milli\meter}. The readout face is the long face at the top of the shape. \textit{Right}: example of a scintillation event in Geant4 from a single X-ray marked as a red dot near the bottom-middle of the teal crystal. The optical pad (brown) and SiPM array (purple) are positioned at the top of the crystal. The paths of the optical photons are depicted by the pink rays, and the hits absorbed by the SiPMs are shown by the red dots along the top.}
\end{figure*}

We are aware of some challenges with using LYSO.
The first is a highly nonlinear light yield at lower energies, with some studies suggesting it can be as low as 55\% below \SI{20}{\kilo\eV}.\cite{Chewpraditkul-2009}
Any nonlinearity in this range would not affect IMPISH due to the strong atmospheric attenuation at tens of \si{\kilo\eV} at balloon altitudes, but it is important to note for applications in future space missions, particularly those with the aim to measure solar thermal emission.
The second, which is of more interest to IMPISH, is its relatively poor intrinsic resolution of 8\% FWHM at \SI{662}{\kilo\eV}, which is almost double the theoretical limit imposed by counting statistics at its brightness.\cite{Chewpraditkul-2009,Gektin-2017}
This performance is sufficiency for IMPISH's science goals, as most of the emission at float altitudes will be nonthermal and wide energy bands will be used for achieving the necessary statistics.
At nonthermal energies, IMPSISH does not need to resolve any spectral features besides a single power law.
However, it must be considered for future space missions that may suffer from cross-contamination between thermal and nonthermal emission, e.g., a mission where it is critical to measure the low energy cut-off in the accelerated electron distribution.
Lastly, LYSO has an intrinsic, low intensity background from \SIrange{100}{1100}{\kilo\eV} due to the radioactive decay of \ce{^{176}Lu} atoms.\cite{Alva-2018}
This could either contaminate measurements or be used as a calibration source depending on the application.
For IMPISH, this will serve as a calibration source.
There is no risk of contamination as the bulk of the emission is at energies outside of our sensitivity and is lower in intensity (on the order of tens to hundreds of X-rays per second) than the solar flares we are trying to observe.
We are currently investigating one alternative scintillator choice, cerium-doped yttrium aluminum perovskite (\ce{YAlO3}:Ce, or YAP:Ce), that would trade some of LYSO's challenges in exchange for others, but we do not have any results yet.
Nevertheless, LYSO is an excellent choice for fast solar spectrometry and is more than capable of achieving the science goals of IMPISH.

\begin{figure*}
   \centering
   \begin{minipage}{\textwidth}
      \centering
      \begin{tabular}{c}
         \includegraphics[width=\linewidth]{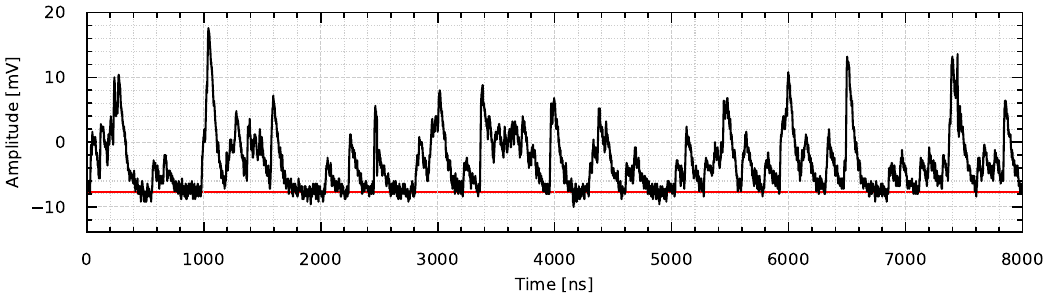}
      \end{tabular}
   \end{minipage}%
   \hfill
   \begin{minipage}{\textwidth}
      \centering
      \begin{tabular}{c}
         \includegraphics[width=\linewidth]{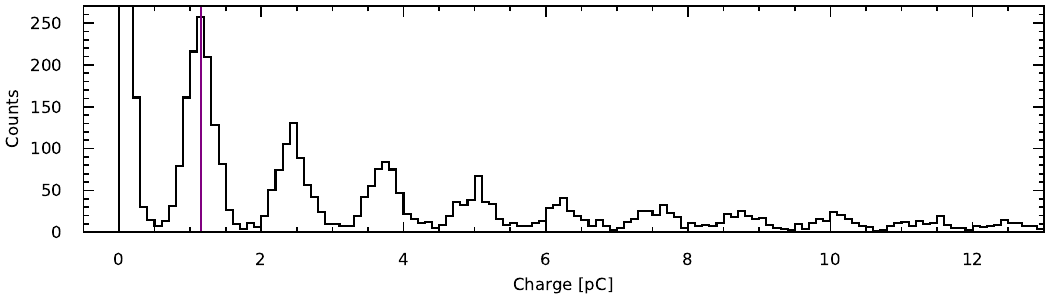}
      \end{tabular}
   \end{minipage}
   \caption
   {\label{fig:dark-counts} \textit{Top}: oscilloscope trace of dark counts measured at a \SI{20}{\mega\hertz} sampling rate from a single SiPM at \SI{12}{\volt} overvoltage with our TIA. The red line denotes our baseline at approximately \SI{-7.7}{\milli\volt}. \textit{Bottom}: spectrum of dark counts in units of charge measured under the same conditions. The voltage pulses were integrated in software to compute their total charge. The bins near \SI{0}{\pico\coulomb} are from electronics noise. The second peak near \SI{1.2}{\pico\coulomb}, marked by the vertical purple lines, are single dark counts and all other peaks are higher order, or integer multiples of (piled up), dark counts.}
\end{figure*}

The stratospheric X-ray environment prompted selecting a crystal geometry that would maximize effective area \textit{and} light collection efficiency (ratio of the amount of light that exited the crystal to the amount that was produced in a scintillation event), while minimizing the number of detector channels.
We performed simulations with Geant4, taking advantage of its X-ray absorption and optical ray tracing models, to guide our choice of geometry and ultimately selected a trapezoidal prism.\cite{Agostinelli-2003}
Figure~\ref{fig:crystal-diagram} shows a sketch of the crystal dimensions and an example of a scintillation event simulated in Geant4.
Having one side of the crystal angled to form the trapezoidal shape introduces an asymmetry that aids in directing the photons towards the readout face at an appropriate angle for transmission.
Further justification through a detailed analysis and exploration of the parameter space, including simulations, crystal geometries, and reflectors, can be found in an accompanying paper.\cite{Oseni-2025}

The reflector around our crystal is 3M's Enhanced Specular Reflector (ESR), which is a non-metallic, polymer-based reflector with $>98\%$ reflectivity across the visible spectrum and excellent overlap with the LYSO emission spectrum.\cite{3M-ESR}
It has the desirable property of being easily cut into arbitrary shapes, either by hand or through the use of precision machines such as laser cutters.
Figure~\ref{fig:lyso-esr-sipm} shows the reflectivity of ESR for photons with a \SI{15}{\degree} incident angle overplotted the LYSO emission spectrum.\cite{Lorincz-2010}
The rapid drop-off in reflectivity at shorter wavelengths is due to fluorescence.\cite{Janecek-2012}
ESR has been demonstrated to be highly specular, even at large incidence angles.\cite{Janecek-2008}

Our chosen SiPM is the Broadcom AFBR-S4N66P014M, which has a \numproduct{6 x 6} \si{\milli\meter\squared} active area, breakdown voltage of \SI{32}{\volt}, and an excellent photon detection efficiency (PDE) that is well-matched to the LYSO emission spectrum with a peak of 63\% at \SI{12}{\volt} overvoltage.
It also has sensitivity to UV wavelengths, which is a rather unique property for a SiPM.
Each IMPISH detector unit will have a total of \num{11} SiPMs to cover the \SI{66}{\milli\meter} readout face of the scintillator crystal.
We will operate all four SiPM arrays at the same bias voltage of about \SI{45}{\volt}, providing a gain of \SI{7.3e6}{\electron\per\photon}.
Each array is coupled to its crystal with optical pads made in-house using the Sylgard 184 silicone elastomer.\cite{Sylgard-184}
Making custom optical pads allows us to match the dimensions of the pad to what is required by our crystal geometry and enables us to test various thicknesses for the pad.\cite{Oseni-2025}

To make the pads, we use \SI{5}{\gram} of part A to \SI{0.5}{\gram} of part B (\num{10}:\num{1} ratio as instructed by the manufacturer) and remove air from the mixture by putting it in a vacuum at about \numrange{1}{10} \si{\milli\bar}.
The mixture is then brought back to atmospheric pressure, and a syringe is used to deposit the mixture onto a glass pane, after which the mixture is once again exposed to vacuum to remove any air introduced during the deposition.
Once the air is removed, the epoxy sits at atmospheric pressure while curing over the course of \numrange{3}{4} days.
This process yields a circular pad with a diameter of about \SI{100}{\milli\meter} and a thickness of about \SI{0.25}{\milli\meter} that we can then cut for use with our detectors.

\begin{figure}[t]
   \begin{center}
      \begin{tabular}{c}
         \includegraphics[width=\linewidth]{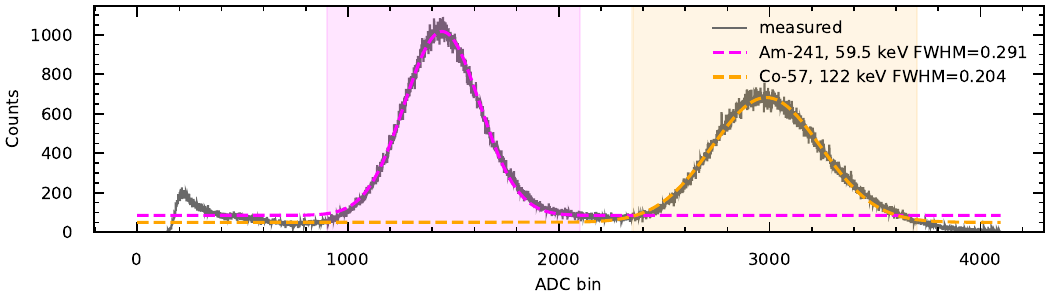}
      \end{tabular}
	\end{center}
   \caption
   {\label{fig:example-spectrum} Combined Am-241 and Co-57 spectra measured with our TIA, shown in gray. The colored, dashed lines are Gaussians fit independently to the two large peaks, which are \SI{59.5}{\kilo\eV} and \SI{122}{\kilo\eV}, with FWHMs of 29\% and 20\%, respectively. The shaded regions denote the fit range for the corresponding Gaussian. The Gaussian fits were also allowed a vertical offset to account for the overlapping peaks.}
\end{figure}

\subsection{Transimpedance Amplifier}%
\label{subsec:transimpedance-amplifier}

We are using a transimpedance amplifier (TIA) configured with a gain of \SI{500}{\ohm} for reading pulses from our SiPM arrays.
Each SiPM array (\num{11} SiPMs) feeds into a single TIA, which is then buffered and sent to the SiPM-3000 for digitization.
A TIA is commonly used with SiPMs since it converts the SiPM photocurrent into a proportional voltage pulse which can be further processed and sampled.
We are using the LTC6268-10 op amp due to its \SI{4}{\giga\hertz} gain bandwidth product to detect the fast LYSO pulses and found that it was \textit{not} necessary to pulse shape after the TIA.

Thermal excitations in the silicon can trigger an avalanche in a SiPM microcell and are indistinguishable from a real photon being detected; these are referred to as dark counts and are well-characterized for any SiPM.
Figure~\ref{fig:dark-counts} demonstrates the sensitivity of our TIA by showing an oscilloscope trace and a spectrum of dark counts obtained from a single SiPM.
In the trace, many dark counts can be seen over the sample of time, some of which are piled up.
The charge per event was calculated by setting a trigger just below the amplitude of a single dark count and integrating the resulting voltage pulses over time.
The spectrum of charge clearly exhibits the dark counts from first-order (single dark count) events to tenth- or eleventh-order.
This data was acquired at \SI{12}{\volt} overvoltage, providing a gain of \SI{7.3e6}{\electron\per\photon}, or about \SI{1.2}{\pico\coulomb}, which is where the first full peak is in this spectrum.

Our specific gain value (\SI{500}{\ohm}) gives IMPISH a dynamic range of \SIrange{\bettersim 10}{240}{\kilo\eV}, depending on the SiPM gain.
This selection was informed by the efficiency of our detector geometry and effects of atmospheric attenuation (e.g. Figure~\ref{fig:atmospheric-attenuation}).\cite{Oseni-2025}
Figure~\ref{fig:example-spectrum} shows a spectrum of Am-241 and Co-57 with our TIA using a single SiPM at \SI{5}{\volt} overvoltage, showing the \SI{59.5}{\kilo\eV} (Am-241) and \SI{122}{\kilo\eV} (Co-57) peaks.
The peaks were independently fit with Gaussians and display the full width at half maximum (FWHM).

\subsection{Custom Bias Supply}%
\label{subsec:custom-bias-supply}

\begin{figure}[t]
   \begin{center}
      \begin{tabular}{c}
         \includegraphics[width=\linewidth]{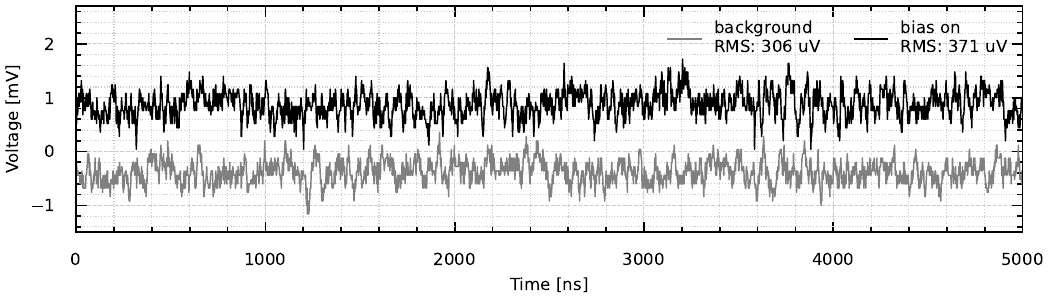}
      \end{tabular}
	\end{center}
   \caption
   {\label{fig:bubba-traces} Sample of traces measured with an oscilloscope. The background trace was recorded with nothing plugged into the oscilloscope, and the ``bias on'' trace was recorded with the bias supply turned on and set to \SI{40}{\volt}. The RMS value is reported for both traces, with ``bias on'' value barely above that of the background and within fluctuations observed on the scope. An artifical DC offset was applied to the traces for clarity.}
   \begin{center}
      \begin{tabular}{c}
         \includegraphics[width=\linewidth]{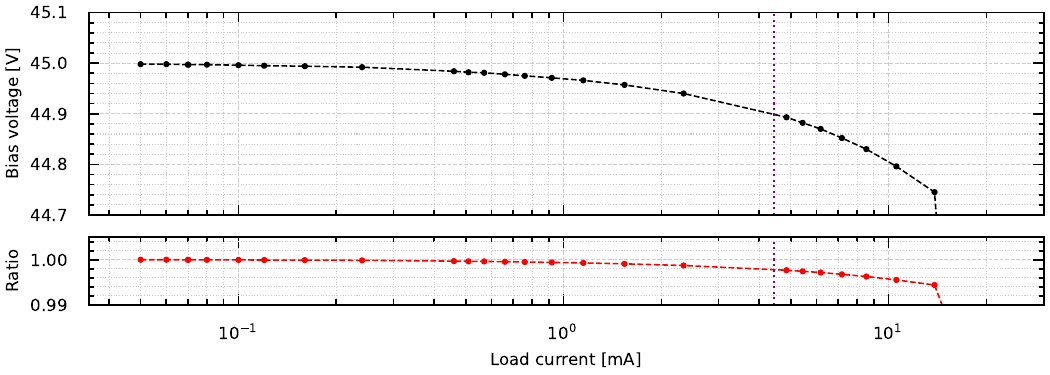}
      \end{tabular}
	\end{center}
   \caption
   {\label{fig:bubba-stiffness} Load regulation for our custom bias supply. The top panel shows how the voltage changes as a function of load current, and the bottom panel shows the ratio of loaded voltage to the initial (nominal) voltage of \SI{45}{\volt}. The purple dotted line in both panels marks a power output of \SI{200}{\milli\watt}, the reported limit for our audio transformer. This amount of current is enough to drive our four detector units.}
\end{figure}

We designed a novel, very low-noise power supply to bias the SiPMs to minimize its contribution to spectral resolution.\footnote{\url{https://github.com/umn-impish/kicad}}
The supply generates a sine wave with an operational amplifier (op amp) low frequency oscillator that is buffered by a power op amp, amplified with an audio transformer, rectified, then regulated down to a stable, DC voltage.
The specific oscillator used is the quad op amp, phase-shift ``Bubba'' oscillator with a frequency of about \SI{2}{\kilo\hertz}.
We chose a frequency much lower than that of the scintillator pulses (on the order of \si{\mega\hertz}) so it did not contaminate the pulses output by the SiPMs.
The use of a linear regulator to maintain the rectified DC voltage helps to further reduce any noise from earlier in the chain and keeps the bias, thus detector response, very stable during operation.
Additionally, four \SI{68}{\micro\farad} aluminum polymer capacitors are used at the output of the rectifier to smooth the ripple and ease the system into and out of periods of higher X-ray flux (i.e. higher current draw).
Aluminum polymer are desirable for space- or near space-based applications due to their large capacitance and benign failure modes.\cite{Liu-2010}
This is in contrast to other constructions, such as a ``wet'' electrolytic or tantalum capacitors.

To demonstrate the very low noise output of this bias design, Figure~\ref{fig:bubba-traces} compares oscilloscope measurements of the bias output to the background.
The background trace was recorded with nothing plugged into the oscilloscope.
The bias was set to \SI{40}{\volt} with no load, and it can be seen that the root mean square (RMS) is only slightly higher (some tens of \si{\micro\volt}) with the bias supply turned on.
This test does \textit{not} demonstrate resilience to noise injected through, for example, electromagnetic interference or switching supply noise on the low voltage power inputs.
Protection against injected noise should be accomplished through proper shielding or filtering appropriate to the characteristics of the anticipated noise.

Our current design requires a $\pm\SI{15}{\volt}$ input with a maximum \SI{60}{\volt} output, which is enough to bias many models of SiPMs, and can drive up to a \SI{200}{\milli\watt} load.
These quantities are determined by the limits of the specific components chosen, but the design itself is very robust and could cater a wide range of inputs and outputs with a different selection of parts.
Notably, our chosen transformer, the Xicon 42TM026-RC, is imposing the \SI{200}{\milli\watt} power load, and the linear regulator, the Analog Devices LT3013MPFE-PBF, is setting the maximum output voltage.

We tested the load regulation of our custom bias supply by measuring how the output voltage and current varied with a resistive load.
Figure~\ref{fig:bubba-stiffness} shows the results of this test, demonstrating great stability (less than half a percent decrease) up to about \SI{10}{\milli\ampere} of load current, which is more than enough to drive our four detector units.
For example, consider the X5 photon rates shown in Figure~\ref{fig:atmospheric-attenuation}.
We can compute an upper limit by multiplying the X-ray rate shown on the plot (\SI{103e5}{\photon\per\second}) by the LYSO light yield (\SI{30}{\photon\per\keV} ignoring nonlinearity) and the peak SiPM PDE of 63\% to obtain approximately \num{1.9e6} optical photons per second.
At our selected overvoltage, a single SiPM produces about \num{4.4e6} dark counts per second (at room temperature), resulting in a reported dark current of \SI{8.6}{\micro\ampere}, implying the total current draw due to the X5 flare will be on the order of \si{\micro\ampere}.
This estimate assumes all the scintillated light exits the crystal and hits the SiPMs, which is not true and will vary depending on the crystal geometry.\cite{Oseni-2025}

An initial vacuum test down to \SI{0.4}{\milli\bar} has shown that the supply remains relatively cool during operation.
Given an \SI{87}{\milli\watt} output load, the warmest component was the power op amp which stabilized at about \SI{48}{\degreeCelsius}.
The parameters of this test were beyond the extremes of what would be encountered during flight (i.e. lower atmospheric pressure than at float altitude, very high power load compared to that demanded by SiPMs, no heat sinking), and the supply performed well within operational limits, showing promise for further, more rigorous vacuum tests and for flight.

\section{Current investigations}%
\label{sec:current-investigations}

This section outlines some ongoing investigations relevant to IMPISH development.
Namely, this includes a description of our preliminary thermal simulations for characterizing the heat dissipation during ascent and at altitude.
As previously mentioned, we are working to understand the properties of our collimator (Section~\ref{subsec:tunsten-collimator}) and also considering an alternative scintillator option (YAP:Ce), but we currently do not have enough results to present at this time.

\subsection{Thermal Simulations}%
\label{subsec:thermal-simulations}

During ascent, ambient temperatures reach as low as \SI{-50}{\degreeCelsius}, which is outside the operating temperature of most electronics.
Indeed, our flight computer is rated only to a minimum of \SI{-20}{\degreeCelsius}.
This poses a threat to the survivability of the payload, so heaters will be used to warm the interior of IMPISH.
SMASH, a previous piggyback instrument on GRIPS-1, used two heaters for a total of \SI{8}{\watt} and found that a higher wattage would have been more comfortable, as their systems reached as low as \SI{-20}{\degreeCelsius} during ascent.\cite{Caspi-2016}
At float altitudes, ambient temperatures are about \SI{-20}{\degreeCelsius}, but atmospheric pressure is very low (approximately 0.1\% of that at sea level) meaning convection will be negligible.
Heaters will likely not be needed at float, but we may need passive measures such as heat sinks or heat straps to maintain cooler temperatures for some components.

Simulations using Thermal Desktop are being designed in order to evaluate the placement and wattage needed for heater(s) during ascent and the exterior coatings needed to stay cool at float altitudes.
It is likely some or all of the outside surfaces will be covered in silverized Teflon tape in order to keep IMPISH cool at float altitudes.
Silverized Teflon has high infrared emissivity and low solar absorptivity and is often used for balloon-borne missions (e.g. SMASH from GRIPS-1).\cite{Kobayashi-2008,Caspi-2016}
However, the necessity of this will be further investigated.
An additional benefit to performing these simulations is that they inform us of temperatures we may expect for our SiPMs.
SiPM properties, particularly the dark count rate and gain, are highly variable with temperature.
A lower operating temperature results in fewer dark counts and thus better performance at lower X-ray energies; the dark count rate is halved with every \SI{10}{\degreeCelsius} decrease.
Similarly, the gain is positively correlated with temperature and can change by about 10\% across a \SI{40}{\degreeCelsius} temperature range.

\section{Conclusions}%
\label{sec:conclusions}

Neither previously nor currently existing instrumentation has been incapable of performing fast cadence solar HXR measurements without significant issues, such as pileup.
RHESSI was limited to a \SI{\geq 2}{\second} cadence without risking data integrity, and STIX is only capable of exploring variations on the order of \SIrange{1}{10}{\second} with its minimum \SI{0.1}{\second} sampling.
Other, non-solar dedicated instruments, such as Fermi GBM, can probe sub-second variations in solar HXRs only with considerable pileup.

The need for a low-cost, low-latency spectrometer is emphasized with the growing demand for space weather forecasting and nowcasting.
Existing solar HXR missions are designed for more complex science objectives, resulting in a distinct lack of instruments monitoring solar flares with HXRs.

IMPISH will be a fast HXR spectrometer designed to explore electron acceleration mechanisms in solar flares.
It is currently designed to operate on a stratospheric balloon, but the intent is to space-qualify the design for use onboard future space-based missions.
The design is cost effective, making use of off-the-shelf components with the use of custom boards, and can easily be adapted to other near-space or space environments.

\acknowledgments
 
We acknowledge and thank NASA's Low Cost Access to Space grant (80NSSC24M0030), NASA's FINESST support (80NSSC23K1621), and the GRIPS platform (80NSSC21K0812).
This work made us of the University of Colorado Space Weather Technology, Research and Education Center's (SWx TREC) \texttt{pymsis} wrapper for atmospheric modeling.
We also thank Dr. Okumura Akira for providing us samples of UV-enhanced ESR for testing in our investigation into using YAP:Ce scintillators.

\bibliography{references}
\bibliographystyle{spiebib}

\end{document}